\documentclass[letterpaper, 10 pt, conference]{ifacconf}

\usepackage{graphicx}      
\usepackage{amsfonts}   
\usepackage{amsmath}   
\usepackage{natbib}        

\usepackage[utf8]{inputenc}
\usepackage[hyphens]{url}
\usepackage{subcaption}
\graphicspath{ {./images/} }

\usepackage{newunicodechar}
\begin{document}
\begin{frontmatter}

\title{Residual Policy Learning for Powertrain Control}

\author[First]{Lindsey Kerbel} 
\author[First]{Beshah Ayalew} 
\author[Second]{Andrej Ivanco}
\author[Second]{Keith Loiselle}

\address[First]{Department of Automotive Engineering, Clemson University, Greenville, SC 29607, USA (e-mail:(lsutto2, beshah)@clemson.edu).}
\address[Second]{Allison Transmission Inc., One Allison Way, Indianapolis, IN, 46222, USA (e-mail:(andrej.ivanco, keith.loiselle)@allisontransmission.com)}
\begin{abstract}                
Eco-driving strategies have been shown to provide significant reductions in fuel consumption. This paper outlines an active driver assistance approach that uses a residual policy learning (RPL) agent trained to provide residual actions to default power train controllers while balancing fuel consumption against other driver-accommodation objectives. Using previous experiences, our RPL agent learns improved traction torque and gear shifting residual policies to adapt the operation of the powertrain to variations and uncertainties in the environment. For comparison, we consider a traditional reinforcement learning (RL) agent trained from scratch. Both agents employ the off-policy Maximum A Posteriori Policy Optimization algorithm with an actor-critic architecture. By implementing on a simulated commercial vehicle in various car-following scenarios, we find that the RPL agent quickly learns significantly improved policies compared to a baseline source policy but in some measures not as good as those eventually possible with the RL agent trained from scratch.

Copyright © 2022 The Authors. This is an open access article under the CC BY-NC-ND license
(https://creativecommons.org/licenses/by-nc-nd/4.0/)

\end{abstract}
\end{frontmatter}

\section{Introduction}
With $29\%$ of greenhouse emissions coming from the transportation sector in the United States, new governmental regulations and incentives are being continuously adopted ~\citep{EmSector} to motivate vehicle original equipment manufacturers (OEMs) to apply various technical solutions in reducing energy usage and emissions. Several measures that attempt to influence the driver's behavior to reduce energy consumption, collectively known as Eco-driving strategies, have been proposed to work with highly engineered powertrains including conventional, hybrid, and electric vehicles~\citep{ecodriv}.
 
Many of these Eco-driving solutions involving passive driver advisory tend to rely on driver compliance and training. Implementing these strategies through active control systems has shown to offer an significant energy consumption savings while mitigating the need to rely on the driver. These solutions use optimal control problem formulations to optimize the velocity trajectory, modulate kinetic energy, and constrain accelerations and are often integrated into an adaptive cruise control (ACC) system~\citep{NieACCMPC}. In~\cite{Yoon}, one approach is proposed to optimize traction torque while still accommodating the driver's request to manage the kinetic energy of a commercial vehicle by using radar information about the preceding vehicle demonstrating a $12\%$ fuel savings.  

However, these control methods rely heavily on powertrain and vehicle models which may be inaccurate in the face of unknown traffic, vehicle conditions and environmental variables. Common issues with model-based controls is the requirement of knowing real-time vehicle and road parameters such as rolling/aerodynamic resistances and vehicle mass, for example.~\cite{realworld} showed that the fuel consumption varies up to $11\%$ due to regional driving patterns in addition to other factors such as vehicle characteristics, weather, road and traffic conditions, and driver tendencies. Despite extensive calibration efforts, control systems deployed with production vehicles cannot anticipate all the variations found in real-world driving scenarios. These systems are often designed to encompass a broad set of drive cycles, and this leads to compromises on the achievable performance for a vehicle dedicated to a specific vocation/fleet.

In a previous work~\citep{kerbel}, we presented a data-driven reinforcement learning (RL) based controller where a vehicle model is not utilized for the control strategy implementation. RL is fundamentally about learning the optimal policy from interactions of the RL agent with the environment without explicit needs for modeling. There are indeed several recent applications of RL to vehicle control including automated lane change maneuvers~\citep{wang2018}, fuel optimal velocity trajectories~\citep{HuPractical}, and optimal transmission gear control~\citep{StepGear}, all of which are integrated into ACC. In most of these cases, including our prior work, the RL agent is trained from scratch. It typically takes several training episodes before a convergent policy is found. 

In this work, we build upon the idea of using an RL-based controller to assist a driver with optimal powertrain control to improve fuel economy. However, instead of learning policies from scratch, we set up the RL agent to learn residual policies that adapt to the actual vocation of the specific vehicle. The starting policy in this framework is the original powertrain control policy shipped with the vehicle (which we call the source policy). This approach is known as residual policy learning (RPL), a method to utilize RL to continuously improve upon a predetermined source policy in an uncertain and dynamic environment~\citep{rpl}.~\cite{rplrobots} and ~\cite{rplshared} demonstrated the potential of this approach to accelerate training of the RL agent by boosting data efficiency where pre-designed complex feedback laws are available as source policies for robots in a manufacturing environment. 

The contribution of this paper is to demonstrate the application of the RPL approach to driver-assist vehicle powertrain control using baseline (OEM equivalent) engineered control policies as source policies. The detailed formulation of the RPL agent is offered and its learning and ultimate performance is compared against that of an RL agent trained from scratch. In this application, we configure the agents to compute both transmission gear selection and traction torque commands.

The rest of the paper is organized as follows. Section 2 outlines the problem formulation. Section 3 details the proposed RPL framework. Section 4 presents results and discussions and Section 5 concludes the paper. 

\section{Problem Formulation}
In this work, we consider a commercial vehicle with a powertrain consisting of a multispeed transmission and an internal combustion engine where the goal is to provide driver assistance via powertrain control. To formulate the driver assistance controller as an RL agent, we model the environment for the Markov Decision Process (MDP) to consist of a driver controlling the ego-vehicle while following a leading vehicle. The state of the environment is then defined as $s= [V_e, A_e, A_{des}, n_g]$, where  $V_e$ is the ego vehicle’s velocity, $A_e$ is the ego vehicle’s acceleration, $A_{des}$ is the driver’s desired input, and $n_g$ is the transmission gear (range).  The actions $a$ consist of the desired traction torque and requested gear change ($T$ and $u_{g}$) where $u_g \epsilon [-1, 0, 1]$ (downshift, remain in current gear, or upshift).

We seek a method to utilize the original powertrain control policy as delivered by the OEM while learning an optimal residual policy and state-action value through repeated interactions with the prevailing environment experienced by the vehicle. To formulate the MDP for this RPL agent, we consider the same environment model as described above but we augment the state vector with actions requested by the default source policies ($T_s$ and $u_{g,s}$). The distinction between the formulations of the regular RL agent and the RPL agent is depicted in Figure 1.
\begin{figure}
\begin{center}
\begin{subfigure}{3.6cm}
  \centering
  \includegraphics[width=3.6cm]{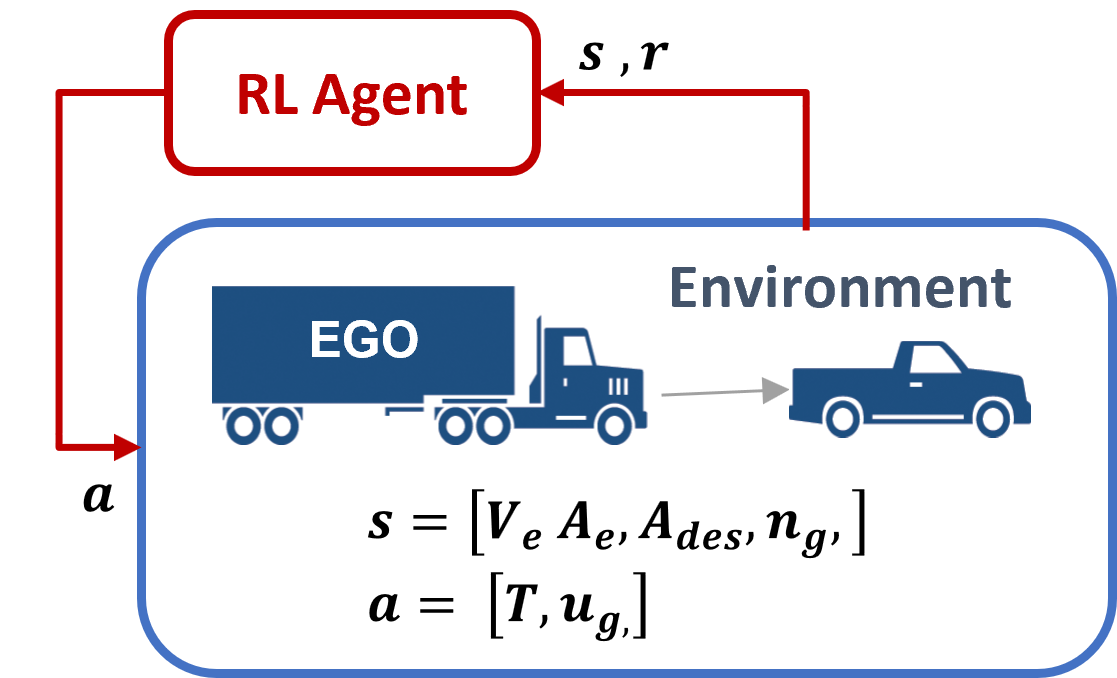} 
  \caption{RL Agent}
  \label{fig:sub1}
\end{subfigure}%
\begin{subfigure}{4.5cm}
  \centering
  \includegraphics[width=4.1cm]{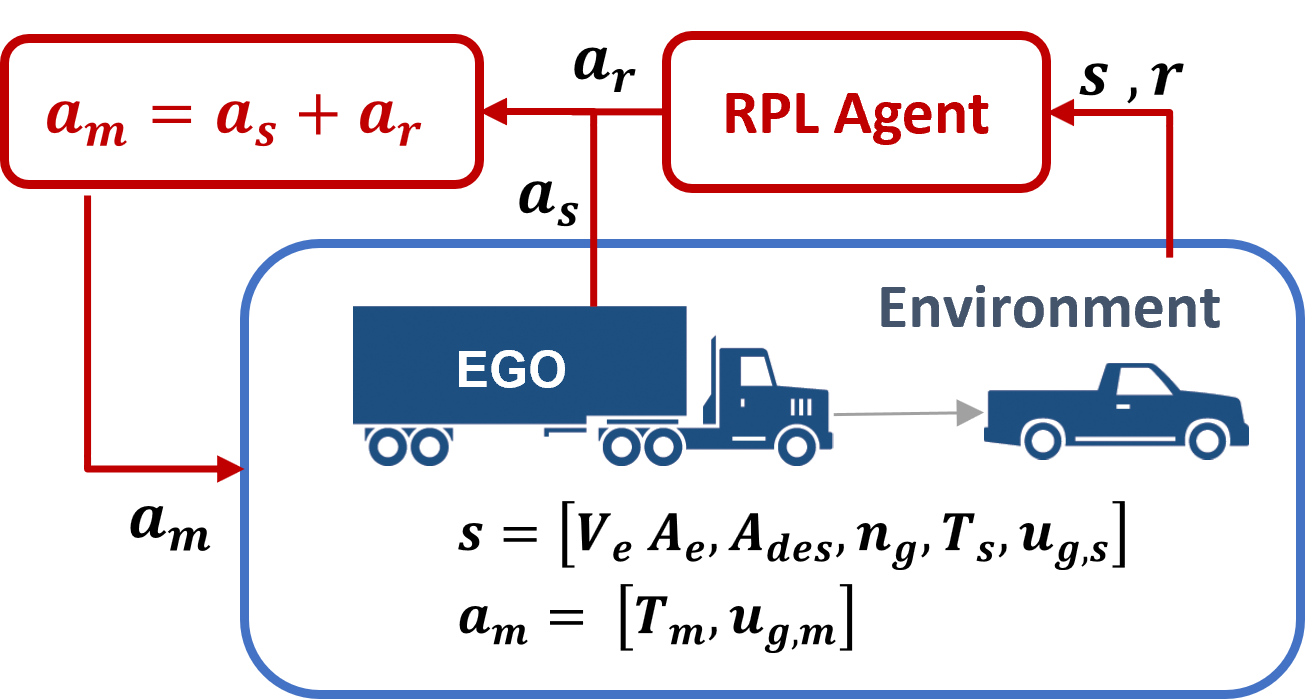} 
  \caption{RPL Agent}
  \label{fig:sub1}
\end{subfigure}%
\caption{Formulation of the driver-assistance controllers with a regular RL agent vs an RPL agent.} 
\label{fig:environment}
\end{center}
\end{figure}

For the RPL agent, the deterministic source actions are summed with actions sampled from learned residual actions $a_r = [T_r, u_{g,r}]$. The action vector applied to the environment is then considered the mixed control action vector $a_m = [T_m, u_{g,m}]$ solved as (1).
\begin{equation} \label{eq1}
a_m = a_s + a_r
\end{equation}
The next state ($s'$) of the environment and the rewards ($r$) are calculated from the current state and the applied mixed action through a detailed vehicle model simulation. In practice, both the state transitions and rewards would be generated from on-board data, often available as on-board diagnostics information, or from telematic and edge devices that log suitably decimated data.

For both of our RL and RPL agents, we define a multi-objective reward function that balances the driver-accommodation and fuel consumption reduction goals expressed through several terms. We seek to minimize the absolute acceleration error with respect to the driver's desired acceleration ($|A_{des} - A_{e}|$) while also minimizing the fuel rate ($\dot{m_f}$) and the traction torque ($T$) magnitude. Further, we penalize gear shifting frequency for driver comfort. Another driveability component is to ensure there is adequate acceleration capability available from the engine to the driver by considering a power reserve ($P_r$) term. Each of these reward components are normalized with respect to their corresponding maximum values to simplify interpretations of weight selections as shown below.
\begin{equation} \label{eq2}
\begin{split}
r(s_t, a_t) =& -W_A\frac{|A_{des_t} - A_{e_{t+1}}|}{\Delta A_{max}} - W_{T}\frac{|T_t|}{T_{max}}- W_{fr}\frac{\dot{m}_{f_{t+1}}}{\dot{m}_{f,max}} \\
& - W_g|n_{g_{t+1}} - n_{g_t}| - W_{pr}\frac{P_{r, max_{t+1}} - P_{r_{t+1}}}{P_{r, max}}
\end{split}
\end{equation}
where $\Delta A_{max}$, ${T_{max}}$, and $\dot{m}_{f,max}$ are the maximum acceleration error, allowed desired torque, and fuel rate. The maximum power reserve $P_{r, max}$ is a function of velocity and is computed from the maximum engine torque curve in each gear~\citep{shiftstrategies}. The weight components $W_A, W_T, W_{fr}, W_g$, and $W_{pr}$ trade-off objectives of minimizing the error in the desired acceleration, traction torque, fuel rate, gear shift frequency, and maintaining available acceleration. For the comparative studies we present later, the reward function remains the same between the RPL and the RL only versions, as they have the same Eco-driving objectives.

\section{Residual Policy Learning Framework}
Figure 2 shows a schematic of the learning framework for the RPL agent. The solid lines illustrate the flow of states and actions between the environment and the controller implemented at each step and the dashed lines represent the information sent to a replay buffer $(s, s', a_r, r)$, where the RPL algorithm samples previous experiences to learn the residual policy.  
\begin{figure}
\begin{center}
\includegraphics[width=8.0cm]{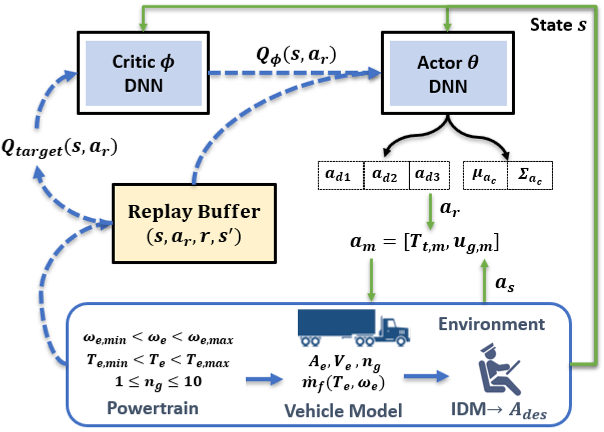}    
\caption{Proposed Residual Policy Learning Architecture} 
\label{fig:architecture}
\end{center}
\end{figure}
We use an actor-critic architecture with a deep neural network for the action policy (actor, parameterized by $\theta$, denoted $\pi_{\theta}$) along with a critic network to approximate a state-action-value function (Q-value, parameterized by $\phi$, denoted $Q_{\phi}(s,a)$).  In the overall framework, the actor approximates the optimal residual actions given the current state to modify the source actions sent to the vehicle system (in the environment).  Through the interactions with the mixed actions taken ($a_m$), the environment generates the next state and calculates a reward.  The critic uses the reward to approximate a value for the state-action pair that was implemented in the environment.  In this case, we can consider the Q-value approximation for the state and residual action pair because the source actions are considered as states.

Since we have a hybrid action space, we train the actor network to approximate both the optimal continuous action (desired torque, $T$) and the discrete action (desired gear change, $u_g$). In our experiments, these actions share the same network parameters with different output layer activation functions. The continuous residual torque action is considered as a Guassian stochastic policy with the mean (Tanh function) and a standard deviation (sigmoid function) as its parameters. The discrete residual gear action is modeled as a categorical distribution through a softmax activation function with three discrete action choices.  We factorize the policy into continuous and discrete components by considering them as independent as described in~\cite{neunert2020}. While this holds true in the practical arrangement of the powertrain control where one can more or less manipulate gear and torque through independent actuation, the independence assumption allows a ready factorization of the joint distribution of the torque and gear actions.

The mixed torque and gear actions are then passed to the powertrain simulation which enforces constraints consistent with the corresponding components. Specifically, the computed mixed torque action request is saturated by the limitations of the engine torque and/or the braking system. Similarly the requested mixed gear action (sum of transmission controller output and the residual gear change request) is limited to one gear change in either direction and the final gear output is enforced by the powertrain engine speed limits at the given vehicle speed. The final available torque and gear actions are applied to the vehicle model to transition to the next state. The environment generates reward signals that encompass the fuel rate ($\dot{m}_f$), the vehicle acceleration ($A_e$), and gear ($n_g$) that occurred as a result of the actions currently applied. These signals are used for computing the total weighted reward according to (2) that we aim to maximize.

In the next section, we briefly describe the reinforcement learning algorithms we use to train the actor-critic setup for the RPL agent as well as the RL only agent.
\section{Reinforcement Learning Algorithm}
We have considered several state-of-the-art on-policy and off-policy algorithms to train our actor-critic networks. In general, off-policy algorithms such as DDPG~\citep{Lillicrap2016ddpg} and SAC~\citep{sac} are known to offer better sample efficiencies, especially in high dimensional and continuous spaces.  Although the RPL architecture can be implemented alongside any RL algorithm, a novel off-policy algorithm, known as maximum posteriori optimization (MPO), was chosen considering that it combines the benefits of on-policy and off-policy algorithms by drawing from a probabilistic inference perspective to optimal control~\citep{abdolmaleki2018}. In our experiments, this algorithm demonstrated stable performance with minimal hyperparameter tuning and did not require special treatment for the hybrid action space other than the one described above. Next, we briefly summarize the key steps of this algorithm as it was implemented for the RPL controller.

The algorithm starts with a policy evaluation step which is done via a critic network that approximates the state-action-value (Q-value) for the policy. To fit the parameters $\phi$ for the critic network, a squared loss function is minimized between the current action value for the current policy iterate $Q_{\theta_k}(s,a|\phi)$ and an estimated target Q-value, $Q_{target}(s_t,a_t)$.  
\begin{equation} \label{eq3}
L(\phi) =  \mathbb{E}\left[Q_{\theta_k}(s_t, a_t|\phi) - Q_{target}(s_t,a_t) \right]^2
\end{equation}
Several algorithms exist to estimate a target Q-value. A stable and efficient method that provides low variance and low bias is desirable. In this work, we adopted a Q-value target estimation method known for stability and efficiency, the Retrace algorithm~\citep{ZhuRetrace}. We refer the reader to this original paper for details of the Retrace algorithm. For now it should suffice that we train the critic network with the Retrace target to obtain the Q-value estimate that minimizes the loss function (3) for the current iterate of the policy.

Given the value function $Q_{\theta_k}(s_t,a_t)$ for the current policy iterate $\pi_{\theta_k}$, the MPO algorithm uses an expectation-maximization scheme to update the policy in two steps~\citep{neunert2020, abdolmaleki2018}. First, with samples from the replay buffer, a non-parametric improved policy $q$ is constructed to maximize $\mathbb{E}_q [Q_{\theta_k}(s_t,a_t)]$ without considerable deviation from the current policy. This optimization problem is posed as: 
\begin{equation} \label{eq4}
\begin{split}
& \max_{q} \mathbb{E}_{q(a|s)} [Q_{\theta_k}(s,a)] \\
& s.t.\; \mathbb{E}_{\mu(s)} \left[ KL \left( q(a|s)|| \pi_{\theta_k}(a|s) \right) \right] < \epsilon
\end{split}
\end{equation}
where $\mu(s)$ is the visitation distribution given in the replay buffer, and the constraint uses a Kullback–Leibler (KL) divergence to keep $q$ close to $\pi_{\theta_k}$ limited by $\epsilon$.

The second step fits a new parametric policy $\pi_{\theta}$ to $q$. The corresponding optimization problem can be written as: 
\begin{equation} \label{eq5}
\theta_{k+1} = \arg\max_{\theta} \mathbb{E}_{s	\sim R} \left[ KL \left(q(a|s) || \pi_{\theta}(a|s) \right) \right] \\
\end{equation}
As mentioned above, given the hybrid action space for the present problem, the KL divergence constraint in (4) is implemented in factorized form in two parts: one for the continuous (torque) policy, and one for the discrete (gear) policy. Equations (3-5) are solved via a gradient-based optimization solver Adam~\citep{Kingma2015AdamAM}. The detailed derivations for the MPO algorithm can be found in~\cite{neunert2020} and ~\cite{abdolmaleki2018}. Further implementation details for the present application can also be found in our straight RL implementation~\citep{kerbel}. Note that the RPL approach is independent of the RL training framework and any state-of-the-art RL algorithms could be applied to learn an optimal residual policy.  

\section{RESULTS AND DISCUSSION}
\subsection{Simulation and RPL Settings}
The simulation environment (including the vehicle model) and RPL algorithm were implemented in Python. A commercial vehicle with a conventional internal combustion engine and 10-speed transmission is modeled to train and demonstrate the RPL framework in car-following scenarios using the dynamics of the vehicle model as shown in (6).
\begin{equation} \label{eq6}
\frac{dV_e}{dt}=\frac{1}{M_{eff}}\left[\frac{T_t}{r_w} +R_r (S_e)+R_a (V_e)+R_g (S_e)\right] 
\end{equation}
where $R_r(S_e)=WC_{r} \cos \psi (S)$, $R_a (V_e)=1/2 \rho C_d A_f V_e^2$, and $R_g (S_e)= W_e \sin \psi (S_e)$. $A_f,C_{1}$, $\rho$, and $\psi$ are the frontal area of the vehicle, the coefficient of rolling resistance, the air density, and the road grade as a function of $S$, respectively. The vehicle’s position, velocity, weight, effective mass, and wheel radius, are listed as $S_e,V_e,W,M_{eff}$ and $r_w$, respectively. The traction torque $T_t$ derived by the powertrain and/or service brakes is the sum of any positive and negative torques applied to the wheels. The fuel rate is interpolated from a table given as a function of engine speed and torque. The engine model limits the torque based on the maximum allowable torque by the engine at the given engine speed. The negative torque is distributed first by applying the maximum engine braking allowed and the remaining torque is applied by the service brake system to the wheels.  

To simulate human-like driving behavior, various driver models have been proposed.  We use a well-established model designed to simulate a typical car-following scenario known as the Intelligent Driver Model (IDM). It uses the relative distance and velocity from the preceding vehicle to calculate a desired acceleration (pedal input) of a driver (for the ego vehicle) in the form $A_{des}$~\citep{TreIDM}. 
 
Although any type of deterministic or feedback controller can be used for the source actions, in this work, the source action for the desired traction torque is a conversion of the driver's desired acceleration $A_{des}$ from the IDM using the longitudinal dynamics as shown in (6). This effectively assumes the source torque policy to ideally invert the demanded acceleration. The baseline also utilizes the desired traction torque directly as calculated for the source torque action. The source gear shifting policy, which is also our baseline shifting strategy, implements a simple instantaneous fuel optimal strategy, where $u_g$ is chosen by minimizing a cost function $J_k = \dot{m}_{f,k} + q_c u_{g,k}$ for the available gears. More detailed information can be found in~\cite{Yoon}. In this case, the source gear controller has full knowledge of the engine's fuel efficiency map whereas the residual action is a learned correction based directly on the environment interactions.

For both the RL and RPL controllers to be compared in the next sub-section, the actor and critic networks of each consist of 3 linear hidden layers (256 units per layer) and the parameters are randomly initialized. The vehicle model and agent parameters are listed in Table 1.
\begin{table}[h]
\caption[width=8.5cm]{Vehicle Model and Hyper-parameters}
\renewcommand{\arraystretch}{1.1}
\label{settings}
\begin{center}
\begin{tabular}{ |cc|cc |}
\hline
\multicolumn{2}{|c|}{Vehicle Model} & \multicolumn{2}{|c|}{Hyper-parameters} \\
\hline\hline
mass & $9070 kg$  & actor learning rate & 5e-5\\
\hline
$A_f$, $C_d$ & $7.71 m^2$, $0.8$   & critic learning rate & 1e-4\\
\hline
 $\Delta t$ & $0.2 s$  & $\gamma$, $\lambda$, $\beta$ & $0.99$, $0.90$, $0.1$\\
\hline
$r_{wheel}$ & $0.498$ & $Q^{retrace}$ steps & $15$\\
\hline 
$A_{max}$ & $2 (m/s)^2$  & KL:$\epsilon_{\mu}, \epsilon_{\sigma}, \epsilon_{d}$ & $0.1, 0.001, 0.1$\\
\hline
$C_r$ & $0.015$  & batch size & $3072$\\
\hline
$t_{headway}$ & $3 s$  & $M$ action samples & $40$\\
\hline
\end{tabular}
\end{center}
\vspace{-2mm}
\end{table}

As mentioned in Figure 1, one distinction between the RPL agent and the RL agent is the augmentation of the source actions to the state vector that allows us to treat the residual policy as an independent policy as a function of the states. It is imperative to ensure the residual policy only enhances the outcome of an already good source policy; that is, it is important to prevent negative transfer. To this end, in our implementation, the residual actions are not combined with the source actions until the critic loss function from (3) falls below a threshold ($\beta$) in its approximation of the Q-value. Thereafter, the residual policy is applied and it then learns as a typical RL algorithm, except with a residual action output and that the state is augmented with the source actions.
\subsection{Analysis and Discussion}
For analysis of the RPL scheme, an RL only agent and an RPL agent were trained equivalently in independent simulations. Both agents neural networks are initialized with randomized parameters for the actor and critic and utilize the same reward weights.  They train the actor/critic networks (learn) by gaining experiences (steps taken in the environment) while following an object vehicle with a drive cycle velocity profile that is adjusted by a constant randomly sampled noise. Every sixty seconds the noise is re-sampled to emulate following a vehicle in randomized traffic on a given route.  For this study, the drive cycle used for training is a randomized version of a combination of the FTP and HUDDS urban drive cycles ~\cite{epaRegs}. The simulated route is repeated with each cycle where the agents learn by randomly sampling a batch of previous experiences every 250 simulation steps (one step consists of an action taken in the dynamic environment).  During the training process, the actions are sampled from the policy's distribution to ensure all actions are explored. As mentioned previously, the residual actions from the RPL agent are not applied until the critic is properly trained whereas the RL agent directly applies its policy actions throughout the entire simulation cycle.     
\begin{figure}[t]
\begin{center}
\includegraphics[width=8.0cm]{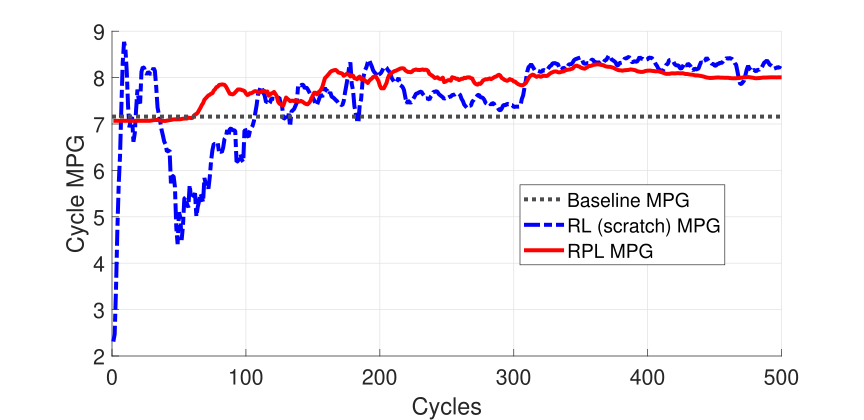}    
\caption{Assessment of fuel economy during learning process} 
\label{fig:episodes}
\end{center}
\end{figure}

To evaluate the learning performance of the two agents, the trained agents are simulated after each intermediate training cycle without noise added to the drive cycle. In these evaluations, the action is taken directly as the highest probability action of the policy. Figure 3 illustrates the evaluated fuel economy measured in miles per gallon (MPG) progression throughout the training cycles. It shows how the RL agent learning from scratch requires numerous cycles before it consistently maintains an improvement in fuel economy while the RPL agent is quick to outperform the source policy. In fact, once the RPL agent is utilized (about $70$ cycles), the fuel economy with the RPL agent is rarely less than the baseline (equivalent to the source) policy's MPG.  This demonstrates that it is possible to train an RPL agent directly on a real vehicle by utilizing the highly developed OEM (as shipped) policies.

After training both agents to convergence, a portion of the evaluation cycle is shown in Figure 4 to demonstrate the actions presented by the baseline (driver only), the RL agent, and the RPL agent.  At this point, the rewards are no longer significantly increasing with additional continued learning cycles. The differences in the final gear and traction torque are shown where both the RL and RPL agents tend to shift to a higher gear earlier and more frequently leading to a generally more optimal engine performance with a similar velocity profile to the baseline. 
\begin{figure}[t]
\begin{center}
\includegraphics[width=8.2cm]{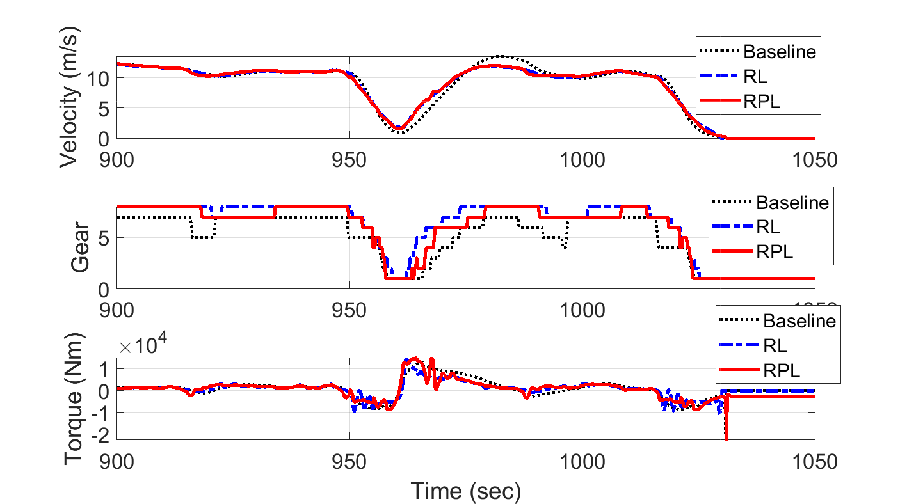}    
\caption{Comparison of RL and RPL Performance} 
\label{fig:episodes}
\end{center}
\end{figure}

A stochastic evaluation with the converged agents is simulated using three standard drive cycles for the leading vehicle including the FTP/HUDDS and two European urban drive cycles: Artemis (Urban and Road), and WLTP~\citep{EUdrivecycles} modified for heavy-duty vehicles.  In this evaluation, noise was added to the velocity profiles to demonstrate the generality of the controllers in various and uncertain environments.  The driver-only baseline, RL only agent, and the RPL agent were simulated 25 times for each of the three stochastic drive cycles. Table 2 summarizes these evaluations using a mean and standard deviation of the cycle fuel consumption (MPG), the root mean square error (RMSE) between the desired and actual acceleration, and the mean number of gear changes applied throughout each of the simulated cycles.  In each case, the overall cycle travel time was minimally affected. 

\begin{table}[h]
\caption{RL and RPL Controller Results}
\vspace{0cm}
\label{results}
\begin{center}
\begin{tabular}{|c|c|c|c|}
\hline
\multicolumn{4}{|c|}{Cycle: FTP/HUDDS} \\ 
\hline\hline 
 & Baseline & RL & RPL \\ 
\hline
MPG & $7.34 \pm 0.11$ & $8.33 \pm 0.16$ & $8.07 \pm 0.21$ \\ 
\hline
$\%$ Difference & $-$  & $+13.45\%$ & $+9.85\%$ \\ 
\hline
Accel RMSE $(m/s^2)$ & $0.36$ & $0.49$ & $0.48$ \\ 
\hline
$\#$ of Shifts & $464$ & $452$ & $610$ \\ 
\hline\hline
\multicolumn{4}{|c|}{Cycle: Artemis Urban} \\
\hline\hline
MPG & $6.35 \pm 0.10$ & $7.25 \pm 0.12$ & $6.74 \pm 0.165$ \\ 
\hline
$\%$ Difference & $-$  & $+14.22\%$ & $+6.10\%$ \\ 
\hline
Accel RMSE $(m/s^2)$  & $0.39$ & $0.58$ & $0.55$ \\ 
\hline
$\#$ of Shifts & $415$ & $460$ & $559$ \\ 
\hline\hline
\multicolumn{4}{|c|}{Cycle: WLTP} \\
\hline\hline
MPG & $7.64 \pm 0.11$ & $8.24 \pm 0.14$ & $8.09 \pm 0.17$ \\ 
\hline
$\%$ Difference & $-$  & $+7.81\%$ & $+5.83\%$ \\ 
\hline
Accel RMSE $(m/s^2)$ & $0.34$& $0.42$ & $0.44$ \\ 
\hline
$\#$ of Shifts & $247$ & $216$ & $291$ \\ 
\hline
\end{tabular}
\end{center}
\vspace{-1mm}
\end{table}

Both RL and RPL agents show a significant improvement in fuel economy while maintaining a similar balance with driver accommodation and travel time. To achieve Eco-driving strategies, the accelerations are limited by the residual traction torque as seen by a slightly higher acceleration error with both agents compared to the baseline. The RL and RPL agents often request a higher gear than the source policy to reduce the engine speed typically leading to a lower fuel rate. The weights in the reward function are adjusted via experimentation to balance these objectives so we can ensure to follow the desired velocity profile (based on the driver's request) and limit over-shifting. The RL agent consistently demonstrates a more fuel beneficial policy in all three drive cycle evaluations as seen in Table 2, but only after training for over $200$ cycles.  

As current RPL work has focused on continuous actions for robots~\citep{multipolar}, to our knowledge, this is the first implementation on a vehicle with a hybrid action space. Directly adding a residual to a continuous action is a natural corrective adjustment as the output of the network has a distribution with a mean value. In the case of the discrete actions, where we simply add the residual action (desired gear shift) to the source action, a bias towards the source policy is seen in the final mixed action.  The action choice is only allowed to modify the source action in one direction as the output of the RPL actor network is a categorical probability of which action to choose.  Summing these probabilities also leads to a bias as the probability of the source action chosen is one (for a deterministic policy), but the probabilities for the residual three action choices sum to one, meaning the source action will always have the highest probability. This explains the difference in the achieved performance of the RPL agent with respect to the RL agent that does not use a source policy.

\section{Conclusion}
In this work, we proposed and demonstrated a novel method that can build on previously engineered powertrain control policies via a residual policy learning agent for active Eco-driving assistance. The RPL agent can learn to adapt the OEM policy shipped with the vehicle to the uncertainties and variations found when driving in the vehicle's actual real-world environment. We observed up to a $10\%$ fuel economy improvement with the RPL policy with respect to a baseline policy that assumes access to information about the vehicle's engine efficiency maps.

The RPL agent was compared to another learning-based controller where the action is taken entirely from an RL agent that is trained from scratch.  The primary benefit of the RPL agent consists of being able to learn from a source policy before its applied to a vehicle, thus allowing it to learn quicker. It improves upon pre-existing knowledge by using interactions from the environment to adapt and adjust the actions for better performance. However, despite the faster learning compared to the RL agent, a bias was observed in the RPL agent's choices towards that of the source policy, which explains why the ultimate performance of the RPL agent still leaves room for improvement. Future research is aimed at how to remove this bias by perhaps some sort of adaptive mixing of the source policy and the residual policy.

\Urlmuskip=0mu plus 1mu\relax
\bibliography{PRL_Powertrain_Control}                                                                
\end{document}